\documentclass[prd,preprint,nofootinbib]{revtex4-1}

\newcount\rem \rem=1
\ifnum \rem=0 
\newcommand{\ygn}[1]{}
\newcommand{\adn}[1]{}
\newcommand{\mgn}[1]{}
\else
\newcommand{\adn}[1]{{\bf \color{blue} [AD:~#1]}}
\newcommand{\mgn}[1]{{\bf \color{orange} [MG:~#1]}}
\fi

\usepackage{amssymb,amsmath}
\usepackage{color}
\usepackage[toc,page]{appendix}
\usepackage{graphicx}
\usepackage{slashed}
\usepackage{tikz}
\usepackage{hyperref}
\usepackage[caption=false]{subfig}

\renewcommand{\Im}{{\cal I}m}
\renewcommand{\Re}{{\cal R}e}

\newcommand{\D}{\mathrm{d}}
\newcommand{\half}{\frac{1}{2}}
\newcommand{\lsim}{\raisebox{-0.13cm}{~\shortstack{$<$ \\[-0.07cm]
      $\sim$}}~}

\newcommand{\ket}[1]{|{#1} \rangle}

\newcommand{\ks}{\ket{K_S}}
\newcommand{\kl}{\ket{K_L}}
\newcommand{\kb}{\ket{\overline{K}{}^0}}

\newcommand{\ko}{\ket{K^0}}

\newcommand{\xtg}{\Delta m t}

\def\beq{\begin{eqnarray}}
\def\eeq{\end{eqnarray}}

\newcounter{mycount}
\newcommand{\pauseen}{\setcounter{mycount}{\value{enumi}}\end{enumerate}}
\newcommand{\resumeen}{\begin{enumerate}\setcounter{enumi}{\value{mycount}}}

\begin{document}
\title{$K\to \mu^{+} \mu^{-}$ beyond the standard model}
\author{Avital Dery}
\email{avital.dery@cornell.edu}
\affiliation{Department of Physics, LEPP, Cornell University, Ithaca, NY 14853, USA}
\author{Mitrajyoti Ghosh}
\email{mg2338@cornell.edu}
\affiliation{Department of Physics, LEPP, Cornell University, Ithaca, NY 14853, USA}

\begin{abstract}
We analyze the New Physics sensitivity of a recently proposed method to measure the CP-violating ${\cal B}(K_S\to\mu^+\mu^-)_{\ell=0}$ decay rate using $K_S - K_L$ interference. We present our findings both in a model-independent EFT approach as well as within several simple NP  scenarios. We discuss the relation with associated observables, most notably ${\cal B}(K_L\to\pi^0\nu\bar\nu)$.
We find that simple NP models can significantly enhance ${\cal B}(K_S\to\mu^+\mu^-)_{\ell=0}$, making this mode a very promising probe of physics beyond the standard model in the kaon sector.
\end{abstract}

\maketitle

\section{Introduction}\label{sec:intro}
%
A recent proposal has demonstrated that short-distance (SD) parameters of the decay $K\to\mu^+\mu^-$ can be cleanly extracted from a measurement of the $K_S-K_L$ interference term in the time dependent rate~\cite{DAmbrosio:2017klp,Dery:2021mct}.
This statement is true to a very good approximation within the SM and any New Physics (NP) model in which the leptonic $(\mu^+\mu^-)$ current is of similar CP structure.
As shown in Ref.~\cite{Dery:2021mct}, once such a measurement is carried out, its results can be interpreted as a measurement of standard model (SM) CKM parameters. Recently, Ref.~\cite{Buras:2021nns} has pointed out that within the SM, the ratio, ${\cal B}(K_S\to\mu^+\mu^-)_{\ell=0} / {\cal B}(K_L\to\pi^0\bar\nu\nu)$ is independent of any SM parameter except for the well-measured $|V_{us}|$ and $m_t$ (and in particular gets rid of any parametric dependence on $|V_{cb}|$). In this work we show that a measurement of ${\cal B}(K_S\to\mu^+\mu^-)_{\ell=0}$ can also serve as a probe of possible NP scenarios.

$K\to\mu^+\mu^-$ is a flavor-changing-neutral-current (FCNC) process, making it a very potent probe of physics beyond the SM, sensitive to high NP scales. The prospects of having a theoretically clean measurement of its parameters are therefore very exciting.
In the following, we investigate the NP reach of this proposed measurement. In other words, we ask the question of to what extent and within which models can the CP-violating mode, ${\cal B}(K_S\to\mu^+\mu^-)_{\ell=0}$, be significantly enhanced compared to the SM.

We first discuss a model-independent generic bound in Sec.~\ref{sec:generic}, we review the basic setup in Sec.~\ref{sec:notation}, then we derive relations to other modes using an EFT approach in Sec.~\ref{sec:EFT}, and analyze specific examples of relevant NP models in Sec.~\ref{sec:NPmodels}. We conclude in Sec.~\ref{sec:conclusion}.

\section{Generic bound}
\label{sec:generic}

The 2020 LHCb bound on $K_S\to\mu^+\mu^-$ reads~\cite{LHCb:2020ycd}
\begin{equation}\label{eq:LHCbbound}
    {\cal B}(K_S\to\mu^+\mu^-) \, < \,  2.1 \, \cdot \, 10^{-10} \, \equiv \, {\cal B}(K_S\to\mu^+\mu^-)_{lim.} \, .
\end{equation}
Within the SM, the prediction for $K_S\to\mu^+\mu^-$ involves a large CP-conserving contribution, dominated by long-distance physics, and a much smaller CP-violating contribution, dominated by short-distance physics. Since these two contributions result in final states of opposite CP, they do not interfere and we have
\begin{equation}
    {\cal B}(K_S\to\mu^+\mu^-) = {\cal B}(K_S\to\mu^+\mu^-)_{\rm CPC}^{(\rm LD)} \, + \, {\cal B}(K_S\to\mu^+\mu^-)^{(\rm SD)}_{\rm CPV} \, .
\end{equation}
The bound of Eq.~(\ref{eq:LHCbbound}) can then be read as a conservative bound on the CP-violating (CPV) short-distance contribution alone,
\begin{equation}
    {\cal B}(K_S\to\mu^+\mu^-)_{\rm CPV} \, < \, {\cal B}(K_S\to\mu^+\mu^-)_{lim.} \, .
\end{equation}
The short-distance CPV contribution to $K_S\to\mu^+\mu^-$ can be identified with the decay of $K_S$ into the CP-odd final state $(\mu^+\mu^-)_{\ell=0}$, where $\ell$ denotes the orbital angular momentum of the dimuon pair. The SM prediction is given by ${\cal B}(K_S\to\mu^+\mu^-)_{\ell=0}^{\rm SM} = 1.64\times 10^{-13}$~\cite{Dery:2021mct}, leaving much room for possible NP contributions,
\begin{equation}\label{eq:boundRmumu}
    R(K_S\to\mu^+\mu^-)_{\ell=0} \, \equiv \,  \frac{{\cal B}(K_S\to\mu^+\mu^-)_{\ell=0}}{{\cal B}(K_S\to\mu^+\mu^-)_{\ell=0}^{\rm SM}} \, \leq \,  \frac{{\cal B}(K_S\to\mu^+\mu^-)_{lim.}}{{\cal B}(K_S\to\mu^+\mu^-)_{\ell=0}^{\rm SM}} \approx 1280 \, .
\end{equation}

As laid out in detail in Ref.~\cite{Dery:2021mct}, within the SM, the decay $K_L\to\mu^+\mu^-$ is CP-conserving, and involves only the $(\mu^+\mu^-)_{\ell=0}$ final state.
In this case we have 
\begin{equation}
    |A(K_L)_{\ell=1}|=0\,,
\end{equation}
which implies that the $K_S - K_L $ interference term involves only $\ell=0$,
\begin{equation}\label{eq:Gammaint}
    \Gamma_{int.} \propto
    \,|A(K_S)_{\ell=0}||A(K_L)_{\ell=0}|\, .
\end{equation}
The observable 
\begin{equation}\label{eq:extractSD}
    \frac{\Gamma_{int.}^2}{{\cal B}(K_L\to\mu^+\mu^-)} \propto |A(K_S)_{\ell=0}|^2,
\end{equation}
then provides a clean measurement of the short-distance, CP-violating parameter $|A(K_S)_{\ell=0}|$.

Any NP that honors the assumption which the analysis of Ref.~\cite{Dery:2021mct} hinges on, that is, keeps $|A(K_L)_{\ell=1}|=0$, retains the form of Eqs.~(\ref{eq:Gammaint}) and~(\ref{eq:extractSD}). 

We can deduce the experimentally allowed range for NP in $\Gamma_{int.}$ from Eqs.~(\ref{eq:boundRmumu}) and (\ref{eq:extractSD}), using the fact that ${\cal B}(K_L\to\mu^+\mu^-)$ is well-measured. We find that
\begin{equation}
    \frac{\Gamma_{int.}}{\Gamma_{int.}^{\rm SM}} = \sqrt{R(K_S\to\mu^+\mu^-)_{\ell=0}} \lsim 36 \, ,
\end{equation}
suggesting that a direct measurement of the $K_S-K_L$ interference term from the time-dependent rate would be sensitive to viable NP scenarios.

\section{Notation and setup}
\label{sec:notation}
%
We use the following standard notation~\cite{ParticleDataGroup:2020ssz}, where the two neutral kaon mass eigenstates, $\ks$ and $\kl$, are linear combinations of the flavor eigenstates:
\beq 
\ket{K_{S}} = p \ko + q \kb,\qquad \ket{K_{L}} = p \ko - q \kb.
\eeq
For the purposes of our analysis, CPV in mixing, which is an ${\cal O}(\varepsilon_K)\sim 10^{-3}$ effect, can be safely neglected and we work in the limit 
\begin{equation}\label{eq:modqp1}
    \left|\frac{q}{p}\right| \, = \, 1\, .
\end{equation}
In this limit, the kaon mass eigenstates are also CP eigenstates, and therefore for final states that have definite CP, the decay amplitude is either purely CP-violating or purely CP-conserving. 
The dimuon final state is a CP eigenstate, which can be in one of two possible configurations of different orbital angular momentum: CP-odd ($\ell =0$) and CP-even ($\ell =1$).

Within the SM, and any extension of it in which the leptonic current inducing the dimuon final state is CP-odd (vectorial, axial-vectorial and pseudoscalar currents all fall under this category), CP-violating short-distance effects only contribute to $(\mu^+\mu^-)_{\ell=0}$. 
In addition, all long-distance contributions are CP-conserving to ${\cal O}(10^{-3})$, see Ref.~\cite{Dery:2021mct} for details.
It follows that in this context:
\begin{enumerate}
    \item Only the $\ell=0$ final state appears in the $K_L$ decay, 
    \begin{equation}
        |A(K_L)_{\ell=1}|=0 \, .
    \end{equation}
    \item A measurement of $K_S-K_L$ interference involves only the $\ell=0$ amplitudes, and is proportional to $A(K_S)_{\ell=0}\times A(K_L)^*_{\ell=0}$.
\end{enumerate}
Thus, as long as NP does not introduce CP-even leptonic operators, the measurement proposed in Ref.~\cite{Dery:2021mct} is a clean measurement of the short-distance amplitude $|A(K_S)_{\ell=0}|$, which is equivalent to a measurement of ${\cal B}(K_S\to\mu^+\mu^-)_{\ell=0}$.

The time dependent decay rate as a function 
of proper time for a neutral kaon beam is given by~\cite{ParticleDataGroup:2020ssz} 
\beq \label{eq:time-dep}
\left(\frac{\D \Gamma}{\D t} \right) = 
{\cal N}_f f(t),
\eeq
where ${\cal N}_f$ is a time-independent normalization factor and 
the function $f(t)$ is given as a sum of four functions
\beq \label{eq:Cdef}
f(t) = C_L e^{-\Gamma_L t}+
C_S \,e^{-\Gamma_S t} + 2 \left[C_{sin} \sin (\xtg) + C_{cos} \cos
(\xtg)\right]e^{-\Gamma t}. 
\eeq
For a pure $K^0$ beam, the coefficients are given by~\cite{Dery:2021mct}
\beq \label{eq:K-Cs}
C_L &=&  |A(K_L)_{\ell=0}|^2, \\ \nonumber
C_S &=& |A(K_S)_{\ell=0}|^2 + \beta_{\mu}^2 |A(K_S)_{\ell=1}|^2 , \\ \nonumber
C_{cos} &=& \Re(A(K_S)_{\ell=0}^*A(K_L)_{\ell=0})=
|A(K_S)_{\ell=0}^*A(K_L)_{\ell=0}|\,\cos\varphi_0,  \\ \nonumber
C_{sin} &=& \Im(A(K_S)_{\ell=0}^*A(K_L)_{\ell=0})=|A(K_S)_{\ell=0}^*A(K_L)_{\ell=0}|
\sin\varphi_0.
\eeq 
where $\varphi_0\equiv\arg(A(K_S)_{\ell=0}^*A(K_L)_{\ell=0})$. The interference effects are embodied by $C_{cos}$ and $C_{sin}$,
\begin{equation}
	\Gamma_{int.} \propto \sqrt{ C_{cos}^2+C_{sin}^2}\, .
\end{equation}

\section{Model-independent analysis using effective operators}
\label{sec:EFT}
%
We consider the effective $|\Delta S|=1$ Hamiltonian,
\begin{eqnarray}
    {\cal H}_{eff.}^{|\Delta S|=1} \, = \, \sum_i C_i O_i,
\end{eqnarray}
where the flavor indices are implicit.
The following six operators are relevant for $K\to\mu^+\mu^-$:
\begin{itemize}
    \item Vectorial operators
    \begin{eqnarray}
        O_{VLL} &=& (\overline Q_L \gamma^\mu Q_L)(\overline L_L \gamma_\mu L_L); \qquad O_{VLR} = (\overline Q_L \gamma^\mu Q_L)(\overline e_R \gamma_\mu e_R), \\ \nonumber
        O_{VRL} &=& (\overline d_R \gamma^\mu d_R)(\overline L_L \gamma_\mu L_L); \,\,\qquad O_{VRR} = (\overline d_R \gamma^\mu d_R)(\overline e_R \gamma_\mu e_R),
    \end{eqnarray}
    
    \item Scalar operators
    \begin{eqnarray}
        O_{SLR} &=& (\overline Q_L d_R)(\overline e_R L_L),  \\ \nonumber
        O_{SRL} &=& (\overline d_R  Q_L)(\overline L_L e_R) \, .
    \end{eqnarray}

\end{itemize}
Since the quark indices we are interested in are non-diagonal, $O_{SLR}$ and $O_{SRL}$ are two distinct operators, not related by hermitian conjugation.
For concreteness, we consider the quark flavor indices to always be $(2,1)$ unless otherwise indicated. 
Note that we do not include tensor operators here, since tensor operators do not contribute to the 2-body decay of a pseudoscalar, as is the case at hand.

We obtain the following general expression for the $K_S\to(\mu^+\mu^-)_{\ell=0}$ rate, in units of the SM expectation,
\begin{eqnarray}\label{eq:generalKS}
    &\,& R(K_S\to\mu^+\mu^-)_{\ell= 0} = \Bigg(1 + \frac{1}{|C_{VLL}^{\rm SM}|\sin\theta_{ct}} \Bigg[A_S \Big( |C_{SLR}^{\rm NP}| \sin \Theta_{SLR}+|C_{SRL}^{\rm NP}| \sin \Theta_{SRL}\Big)  \\
    &\,& \qquad + \, |C_{VLL}^{\rm NP}| \sin \Theta_{VLL}-|C_{VLR}^{\rm NP}| \sin \Theta_{VLR}-|C_{VRL}^{\rm NP}| \sin \Theta_{VRL}+|C_{VRR}^{\rm NP}| \sin \Theta_{VRR} \Bigg]\Bigg)^2 \nonumber\, ,
\end{eqnarray}
where $A_S \equiv \frac{m_K^2/m_\mu} {2(m_s+m_d)}$ is the so-called scalar enhancement factor (see, for example, the discussion around Eq. (28) of Ref.~\cite{Dorsner:2016wpm}), $\Theta_{i}$ is the basis independent phase between the mixing and the Wilson coefficient,
\begin{equation}\label{eq:Theta}
    \Theta_{i} \equiv \half \arg \left(\frac{q}{p}\right) - \arg (C_{i}^{\rm NP})\, ,
\end{equation} 
and~\cite{Dery:2021mct}
\begin{equation}\label{eq:CSMdef}
    |C_{VLL}^{\rm SM}|\sin\theta_{ct} = \left|\frac{G_F}{\sqrt{2}}\frac{2\alpha Y(x_t)}{\pi \sin_W^2}  {\cal I}m\left(-\frac{V_{ts}^*V_{td}}{V_{cs}^*V_{cd}}\right)\right| \, .
\end{equation}
It is important to note, that the scalar operators, $O_{SLR}$ and $O_{SRL}$, induce both the $\ell=0$ and the $\ell=1$ final states, since they include both pseudo-scalar ($P$) and scalar ($S$) leptonic currents. Only the combination $(O_{SRL} + O_{SLR})$ can in general protect the assumption of $|A(K_L)_{\ell=1}| = 0$.

By taking any Wilson coefficient to be ${\cal O}(1/\Lambda^2)$, where $\Lambda$ is the scale of NP, we learn that a measurement of ${\cal B}(K_S\to\mu^+\mu^-)_{\ell=0}$ that saturates the current experimental upper bound would be sensitive to NP scales of up to $\Lambda \sim 40\,{\rm TeV}$ for vectorial operators, and up to $\Lambda \sim 130 \, {\rm TeV}$ for scalar operators.

\subsection{The relation between $K_S\to(\mu^+\mu^-)_{\ell=0}$ and $K_L\to\pi^0\bar\nu\nu$}
Of the six operators, $O_{VLL}$ and $O_{VRL}$ contribute additionally to $K_L\to\pi^0\nu\bar\nu$.
The general expression for $K_L\to\pi^0\bar\nu\nu$, assuming diagonal couplings in flavor space, is given by:
\begin{eqnarray}
   R(K_L \to \pi^0 \bar{\nu} \nu) &=& \frac{1}{3} \sum_{i=e,\mu,\tau} \left(1 + \frac{|(C_{VLL}^{\rm NP})_i| \sin \Theta_{VLL,i}+|(C_{VRL}^{\rm NP})_i| \sin \Theta_{VRL,i}}{|C_{VLL}^{\rm SM}|\sin\theta_{ct}}\right)^2,
\end{eqnarray}
where $R(X)$ denotes the rate of $X$ in units of the SM prediction.
\begin{itemize}
    \item \textbf{Models with lepton-flavor universality.} \\
    $K_S\to(\mu^+\mu^-)_{\ell=0}$ and $K_L\to\pi^0\bar\nu\nu$ are both CPV processes and both arise in the SM from a single operator, ${\cal O}_{VLL}$. This leads to simple relations between their rates, in the case of models that are lepton flavor universal (LFU), and involve only the lepton-doublet vectorial operators, $\{O_{VLL}, O_{VRL} \}$. We have,
    \begin{eqnarray}
         R(K_S\to\mu^+\mu^-)_{\ell = 0}^{\{O_{VLL}, O_{VRL} \}} \,  &=& \, \Bigg(1\, + \, \frac{|C_{VLL}^{\rm NP}|\sin\Theta_{VLL}^{\rm NP} - |C_{VRL}^{\rm NP}|\sin\Theta_{VRL}^{\rm NP}}{|C_{VLL}^{\rm SM}|\sin\theta_{ct}} \Bigg)^2\, , \\ \nonumber
         R(K_L\to\pi^0\bar\nu\nu)^{\rm LFU} \,  &=& \, \Bigg(1 + \frac{|C_{VLL}^{\rm NP}|\sin\Theta_{VLL}^{\rm NP} + |C_{VRL}^{\rm NP}|\sin\Theta_{VRL}^{\rm NP}}{|C_{VLL}^{\rm SM}|\sin\theta_{ct}} \Bigg)^2\, .`
    \end{eqnarray}
    The sign differences in the two expressions result entirely from the fact that the first process is sensitive only to the axial hadronic current, while the second is only sensitive to the vector hadronic current.
    We conclude the following:
    \begin{enumerate}
        \item In models where only $C_{VLL}^{\rm NP}$ is turned on, we have 
        \begin{eqnarray}
            R(K_S\to\mu^+\mu^-)_{\ell=0}^{C_{VLL}} \, &=& \,R(K_L\to\pi^0\bar\nu\nu)^{C_{VLL}}_{\rm LFU} \, .
        \end{eqnarray}
        Then, by using the Grossman--Nir (GN) bound~\cite{Grossman:1997sk} to place an experimental constraint on $R(K_L\to\pi^0\bar\nu\nu)$,
        \begin{eqnarray}\label{eq:GN}
            R(K_L\to\pi^0\bar\nu\nu)^{C_{VLL}}_{\rm LFU}\, \underset{\rm GN}{\lsim} \,  \frac{4.3\cdot {\cal B}(K^+\to\pi^+\bar\nu\nu)}{{\cal B}(K_L\to\pi^0\bar\nu\nu)_{\rm SM}} \, \lsim 26 \, ,
        \end{eqnarray}
        the following bound can be set on the deviation from the SM in ${\cal B}(K_S\to \mu^+\mu^-)_{\ell=0}$, 
       \begin{eqnarray}
            R(K_S\to\mu^+\mu^-)_{\ell=0}^{C_{VLL}} \,\underset{\rm GN}{\lsim} \,  26 \, .
        \end{eqnarray}

        \item In models where only $C_{VRL}^{\rm NP}$ is turned on,
        \begin{eqnarray}
            \frac{R(K_S\to\mu^+\mu^-)_{\ell=0}^{C_{VRL}}}{R(K_L\to\pi^0\bar\nu\nu)^{C_{VRL}}_{\rm LFU}} \, = \, \frac{\left(1 - \frac{|C_{VRL}^{\rm NP}|}{|C_{VLL}^{\rm SM}|} \frac{\sin\Theta_{VRL}^{\rm NP}}{\sin\theta_{ct}}\right)^2}{\left(1 + \frac{|C_{VRL}^{\rm NP}|}{|C_{VLL}^{\rm SM}|} \frac{\sin\Theta_{VRL}^{\rm NP}}{\sin\theta_{ct}}\right)^2} \, .
        \end{eqnarray}
        Then, the GN bound on $R(K_L\to\pi^0\bar\nu\nu)$ results in
        \begin{eqnarray}
            R(K_S\to\mu^+\mu^-)_{\ell=0}^{C_{VRL}} \underset{\rm GN}{\lsim} 50 \, .
        \end{eqnarray}
        \begin{figure}[t!]
        	\includegraphics[width=0.8\textwidth]{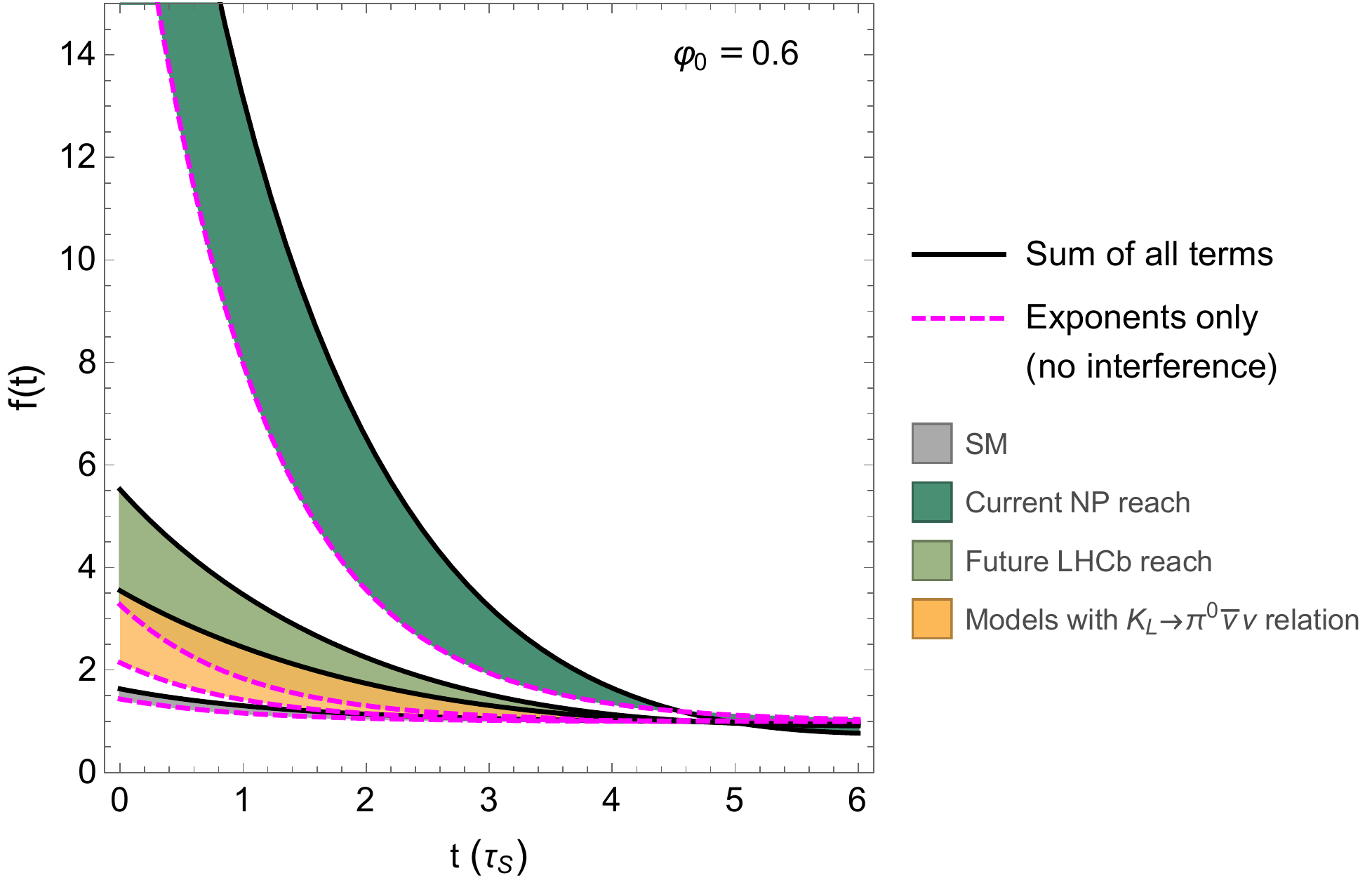}
        	\caption{The time dependence in the decay of a $K^0$ beam into $\mu^+\mu^-$, as given by Eq.~(\ref{eq:time-dep}), for magnitudes of the $K_S\to(\mu^+\mu^-)_{\ell=0}$ amplitude given by: the SM prediction (gray), the current experimental upper bound (dark green), the upper bound for models where the rate is correlated with $K_L\to\pi^0\bar\nu\nu$ (yellow), and the expected future reach at LHCb (light green). The integral between the solid and dashed curves corresponds to the magnitude of interference effects.}
        	\label{fig:NPreach}
        \end{figure} 
        \item Hence, if NP is discovered in a future measurement of $R(K_S\to\mu^+\mu^-)_{\ell=0}$, with a larger value than the above bounds, we will be able to exclude models that turn on only one of $\{O_{VLL}, O_{VRL} \}$.
        
        \item If, on the other hand, a future bound is set on $R(K_S\to\mu^+\mu^-)_{\ell=0}$ that is more stringent than the above, it has the potential to become the leading constraint on $R(K_L\to\pi^0\bar\nu\nu)$ in the framework of a single LFU operator.
        
        \item If more than one operator is turned on, or if any of the operators involving right-handed leptons or scalar currents ($\{{\cal O}_{VLR},\, {\cal O}_{VRR},\, {\cal O}_{SLR},\, {\cal O}_{SRL} \}$) are at play, then the two modes are in general completely independent.
    \end{enumerate}
    
    We note that similar relations are expected to apply between $K_S\to(\mu^+\mu^-)_{\ell=0}$ and the direct CPV contributions in  $K_L\to\pi^0\mu^+\mu^-$ and $K_S\to\pi^0\pi^0\mu^+\mu^-$. These would apply for any single operator that affects $K_S\to(\mu^+\mu^-)_{\ell=0}$. We leave a detailed discussion of these modes to a future work.

    \item \textbf{Models that break lepton-flavor universality.}\\
    In general, there could be operators that contribute to $K_L\to\pi^0\nu\bar\nu$ but not to $K\to\mu^+\mu^-$. These are analogous to $O_{VLL},\,O_{VRL}$ with lepton flavor indices different from $(2,2)$.
    
    We note further that the experimental signature for $K_L\to\pi^0\nu\bar\nu$ involves missing energy, hence it also captures scenarios with exotic undetected particles, to which $K_S\to(\mu^+\mu^-)_{\ell=0}$ is insensitive.
\end{itemize}
The time dependent rate of Eq.~(\ref{eq:time-dep}) for a pure $K^0$ beam, for a choice of the unknown phase $\varphi_0 \equiv \arg(A(K_S)_{\ell=0}^*A(K_L)_{\ell=1})$ is plotted in Fig.~\ref{fig:NPreach}. It is apparent that both the total rate and the effect of interference (depicted by the area between the dashed and solid curves) can be greatly enhanced compared to the SM. The estimate for the future reach of LHCb ${\cal B}(K_S\to\mu^+\mu^-)$ searches is taken from Ref.~\cite{Alves:2018npj}.

\section{Explicit NP models}
\label{sec:NPmodels}
%
In this section we present a few examples of simple models in which ${\cal B}(K_S\to\mu^+\mu^-)_{\ell=0}$ is enhanced compared to the SM expectation, and discuss the relevant constraints in each.
We present three models: two scalar leptoquark representations, each inducing a different vectorial effective operator, and a model with an extra scalar doublet, which introduces the scalar effective operators.
Many other possible models beyond the SM exist that can affect $K\to\mu^+\mu^-$, and some may have unavoidable implications on additional constraints untouched by our choice of  toy models. For example, models with flavor changing $Z$ couplings can enhance ${\cal B}(K_S\to\mu^+\mu^-)_{\ell=0}$, but introduce additional strong constraints from $\varepsilon^\prime / \varepsilon$ that generally restrict the contribution to ${\cal B}(K_S\to\mu^+\mu^-)_{\ell=0}$ to be small. We therefore do not discuss this further here.

\subsection{Scalar Leptoquark : $\tilde S_1(\bar 3, 1, 4/3)$}
\label{sec:LQS1}
As a first example of a simple model that can contribute to $K_S\to(\mu^+\mu^-)_{\ell=0}$, we consider a scalar leptoquark. 
A detailed review of leptoquarks and their phenomenology can be found, for example, in Ref.~\cite{Dorsner:2016wpm}. 
Here we discuss two examples of scalar leptoquark representations which demonstrate some of the characteristic features of the $K\to\mu^+\mu^-$ phenomenology.

We first consider a scalar leptoquark in the following SM gauge group representation,
\begin{equation}
    \tilde S_1 \sim (\bar 3, 1)_{4/3}\, .
\end{equation}
The relevant Lagrangian terms are given by
\begin{equation}
    {\cal L}_{\tilde S_1} \supset g_{12}\, \tilde S_1\overline d_R^C\mu_R + g_{22}\, \tilde S_1\overline s_R^C \mu_R\, + \, h.c.\, ,
\end{equation}
with $\psi^C = C\overline \psi^T,\, C=i\gamma^2\gamma^0$.
After integrating out the leptoquark field, we are left with the following dimension six operator,
\begin{equation}\label{eq:S1tildeO}
    \frac{g_{12}g_{22}^*}{2 M^2_{\tilde S_1}} (\overline d_R^C\mu_R)(\overline \mu_R s_R^C) = \frac{g_{12}g_{22}^*}{4 M^2_{\tilde S_1}} (\overline s_R \gamma^\mu d_R)(\overline \mu_R \gamma_\mu \mu_R),
\end{equation}
where in the last step we used a Fierz transformation. In the language of the effective operators of section~\ref{sec:EFT}, this model induces the operator $O_{VRR}$, with $C_{VRR}^{\rm NP} = \frac{g_{12}g_{22}^*}{4M_{\tilde S_1}^2}$. 

We then have,
\begin{eqnarray}\label{eq:S1tildeR}
    R(K_S\to\mu^+\mu^-)_{\ell=0}^{\tilde S_1}  &=& 
    \left(1\, + \, \frac{\left|g_{12}g_{22}\right|\sin\Theta_{\tilde S_1}}{4 M^2_{\tilde S_1}|C_{VLL}^{\rm SM}| \sin\theta_{ct}} \right)^2\, ,
\end{eqnarray}
where, as in Eq.~(\ref{eq:Theta}), the angle $\Theta_{\tilde S_1}$ is defined as the phase between the Wilson coefficient and the mixing,
\begin{equation}\label{eq:ThetaS1}
    \Theta_{\tilde S_1} \equiv \half \arg \left(\frac{q}{p}\right) - \arg(g_{12}g_{22}^*),
\end{equation}
and  $|C_{VLL}^{\rm SM}|$ is defined in Eq.~(\ref{eq:CSMdef}).
In order to saturate the experimental bound, $R(K_S\to\mu^+\mu^-)_{\ell=0}\approx 1280$, we require
\begin{eqnarray}\label{eq:LQ1req}
    \left|g_{12}g_{22}\sin \Theta_{\tilde S_1}\right| \approx 3.4\cdot 10^{-3} \left(\frac{M_{\tilde S_1}}{{\rm TeV}}\right)^2\, .
\end{eqnarray}

There are several constraints on the model parameters.
Direct searches at ATLAS and CMS for leptoquark states with ${\cal O}(1)$ branching ratios into a muon and a light quark result in lower bounds on the leptoquark mass of~\cite{ATLAS:2021oiz,CMS:2018lab}
\begin{equation}
    M_{\tilde S_1} \gtrsim 1.7\,{\rm TeV} \, .
\end{equation}
Therefore Eq.~(\ref{eq:LQ1req}) can be rewritten as
\begin{equation}\label{eq:boundS1}
    \left|g_{12}g_{22}\sin \Theta_{\tilde S_1}\right| \approx 3.4\cdot 10^{-3} \left(\frac{M_{\tilde S_1}}{{\rm TeV}}\right)^2 \gtrsim 9.3\cdot 10^{-3}\, .
\end{equation}

The same operator of Eq.~(\ref{eq:S1tildeO}) induces $K^0-\overline K{}^0$ mixing via loop diagrams~\cite{Mandal:2019gff}, inducing a contribution to $M_{12}$,
\begin{equation}
    M_{12} = M_{12}^{\rm SM} + \frac{f_K^2 \hat B_K m_K}{384\pi^2 M_{\tilde S_1}^2} (g_{12}^*g_{22})^2.
\end{equation}
We use
\begin{eqnarray}\label{eq:Mixing}
    \Delta m_K = \frac{4{\cal R}e(M_{12}\Gamma_{12}^*)}{\Delta\Gamma}, \qquad |\varepsilon_K| \approx -\frac{{\cal I}m(M_{12}\Gamma_{12}^*)}{\sqrt{2}|M_{12}\Gamma_{12}|}\, ,
\end{eqnarray}
together with the fact that in the kaon system we have $\Gamma_{12}\approx \bar A_0 A_0^*$, where $A_0$ is the decay amplitude of $K^0$ into $(\pi\pi)_{I=0}$. This allows us to identify the relevant physical phase in the NP contribution, and relate it to the physical phase relevant for $K\to\mu\mu$, 
\begin{eqnarray}
    \arg\left(M_{12}^{\rm NP}\Gamma_{12}^*\right) \approx \arg\left(M_{12}^{\rm NP} \bar A_0^* A_0\right) \approx \arg\left(\Big[g_{12}^* g_{22}\Big]^2\frac{q}{p}\right) = -2\Theta_{\tilde S_1}.
\end{eqnarray}
where in the next to last step we used the fact that CPV in $K\to\pi\pi$ is negligible, which is equivalent to neglecting CPV in mixing, as in Eq.~(\ref{eq:modqp1}).

\begin{figure}[t]
 \begin{center}
  \includegraphics[width=0.8\textwidth]{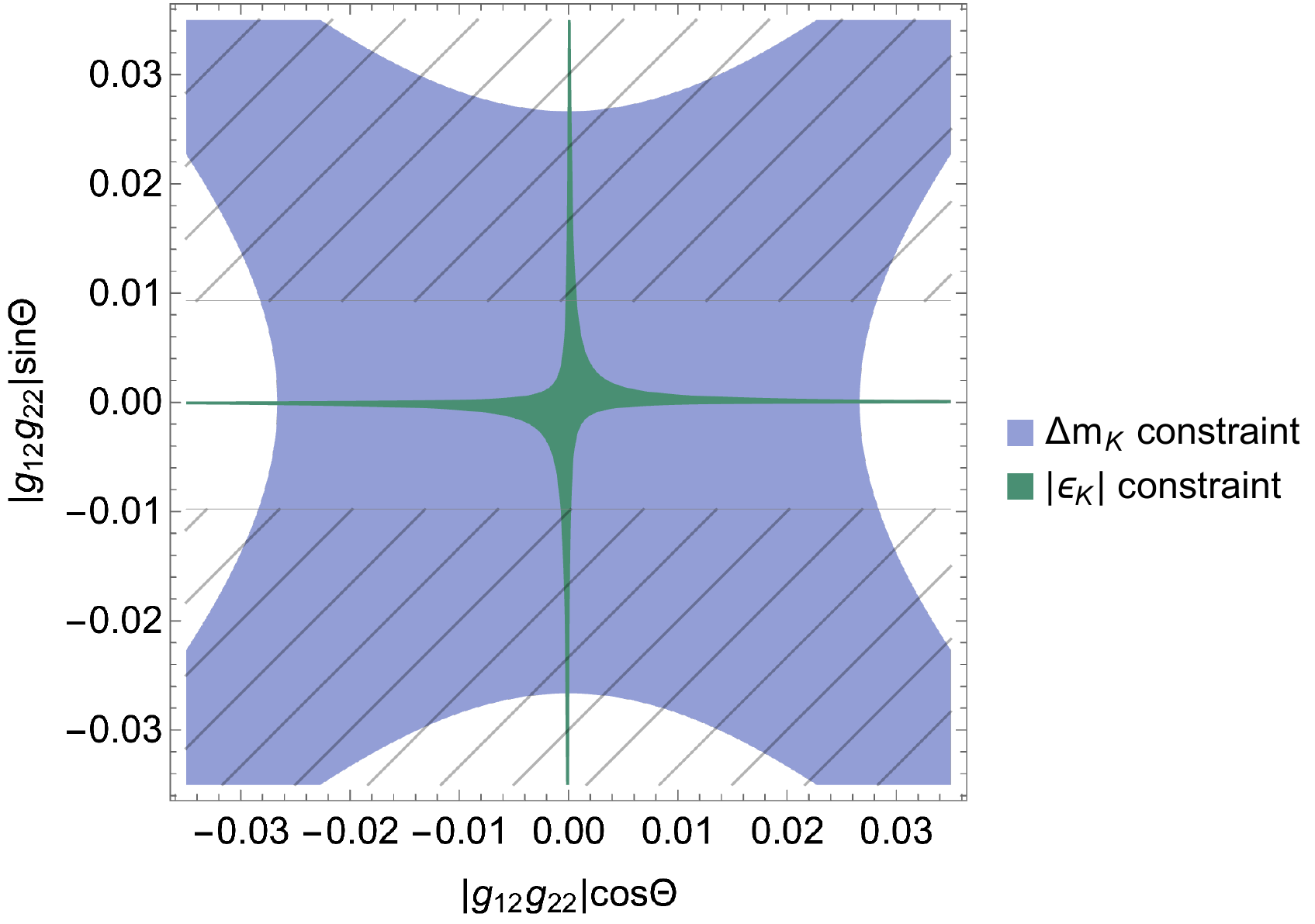}
  \caption{Allowed regions from $K^0-\overline K{}^0$ mixing. The hatched regions are where the bound on ${\cal B}(K_S\to\mu^+\mu^-)_{\ell=0}$ can be saturated (see Eq.~(\ref{eq:boundS1})).}
\label{fig:LQmixing}
\end{center}
\end{figure}

The allowed regions due to mixing are plotted in Fig.~\ref{fig:LQmixing}.
The measurement of $|\varepsilon_K|$ bounds $\left|g_{12}g_{22}\right|^2\sin \Theta_{\tilde S_1}\cos\Theta_{\tilde S_1} $, inducing an inverse relation between the magnitudes of the $\sin\Theta_{\tilde S_1}$ and $\cos\Theta_{\tilde S_1}$, and the constraint from $\Delta m_K$ bounds $\left|g_{12}g_{22}\right|^2\left|\cos^2\Theta_{\tilde S_1} - \sin^2\Theta_{\tilde S_1}\right|$.
The constraint from $|\varepsilon_K|$ ensures that for larger values of $\sin\Theta_{\tilde S_1}$, such that the bound on ${\cal B}(K_S\to\mu^+\mu^-)_{\ell=0}$ is saturated, $\cos\Theta_{\tilde S_1}$ has to be small, such that the contribution to ${\cal B}(K_L\to\mu^+\mu^-)_{\ell=0}$ is well below the theoretical error.
The allowed ranges, such that all constraints are satisfied and the bound on ${\cal B}(K_S\to\mu^+\mu^-)_{\ell=0}$ is saturated, can be summarized as
\begin{eqnarray}
    |g_{12}g_{22}| \gtrsim 9.3\cdot 10^{-3} \qquad \text{AND} \qquad |\cos\Theta_{\tilde S_1}| \lsim 0.08 \, .
\end{eqnarray}

We deduce that the $\tilde S_1$ model can saturate the experimental bound of $R(K_S\to\mu^+\mu^-)_{\ell=0}\leq 1.3\times 10^3$, without violating the constraints from mixing and direct searches.
The interference term could be enhanced by ${\cal O}(30)$ compared to the SM within this model.

\subsection{$S_3(\bar 3,3,1/3)$}
Next we consider an ${\rm SU}(2)$ triplet leptoquark,
\begin{equation}
    S_3\, \sim \, (\bar 3, 3)_{1/3}\, .
\end{equation}
The relevant Lagrangian term is 
\begin{equation}
    {\cal L}_{S_3} \supset g^{QL}\,  (\overline Q_L^C)^a \epsilon_{ab} (\tau_i S_3^i)^{bc}  (L_L)_c \, + \, h.c.\, ,
\end{equation}
where $\tau_i$ are pauli matrices in SU$(2)_L$ space.

After EWSB, we have 
\begin{eqnarray}
    {\cal L}_{S_3} &\supset & g^{QL}_{12}\, \Bigg[\overline d_L^C \bigg(S_3^{(4/3)}  \mu_L + S_3^{(1/3)} {\nu_\mu}_L\bigg) + (\overline u_L^c)_i V_{id} \bigg(S_3^{(1/3)}\mu_L - S_3^{(-2/3)} {\nu_\mu}_L\bigg)\Bigg]  \\ \nonumber
    &+& g^{QL}_{22}\, \Bigg[\overline s_L^C \bigg(S_3^{(4/3)}  \mu_L + S_3^{(1/3)} {\nu_\mu}_L\bigg) + (\overline u_L^c)_i V_{si} \bigg(S_3^{(1/3)}\mu_L - S_3^{(-2/3)} {\nu_\mu}_L\bigg)\Bigg]\, + \, h.c.\, ,
\end{eqnarray}
where $(\overline u_L)_i = (\bar u_L, \bar c_L, \bar t_L)$, and $V$ is the CKM matrix.
Integrating out the leptoquark states, we are left with a list of effective 4-fermion operators. Mediating $d\to s$ transitions, we have:
\begin{eqnarray}
    \frac{g^{QL}_{12}g^{QL*}_{22}}{2 M^2_{S_3^{(4/3)}}} (\overline d_L^C\mu_L)(\overline \mu_L s_L^C)  &+& \frac{g^{QL}_{12}g^{QL*}_{22}}{2 M^2_{S_3^{(1/3)}}} (\overline d_L^C {\nu_\mu}_L)(\overline {\nu_\mu}_L s_L^C) \\ \nonumber
    &=& \frac{g^{QL}_{12}g^{QL}_{22*}}{4 M^2_{S_3^{(4/3)}}} (\overline s_L \gamma^\mu d_L)(\overline \mu_L \gamma_\mu \mu_L) + \frac{g^{QL}_{12}g^{QL*}_{22}}{4 M^2_{S_3^{(1/3)}}} (\overline s_L \gamma^\mu d_L)(\overline {\nu_\mu}_L \gamma_\mu {\nu_\mu}_L).
\end{eqnarray}
There are also $u_i\to u_j, \,u_i\to s,$ and $u_i\to d$ transitions induced in this model, but they do not introduce relevant bounds.

In the $SU(2)_L$ limit, $M_{S_3}^{(4/3)} = M_{S_3}^{(1/3)} $, the effective operator generated by this model is $O_{VLL}$, with $C_{VLL}^{\rm NP} = \frac{g_{12}^{QL}g_{22}^{QL*}}{4M_{S_3}^2}$. Therefore, as expected, the limit on $R(K_L\to\pi^0\bar\nu\nu)$ (Eq.~(\ref{eq:GN})) translates into a limit on $R(K_S\to\mu^+\mu^-)_{\ell=0}$,
\begin{eqnarray}
    R(K_S\to\mu^+\mu^-)_{\ell=0} \, = \, R(K_L\to\pi^0\bar\nu\nu) = \left(1 \, + \, \frac{|g_{12}^{QL}{g_{22}^{QL}}|\sin\Theta_{S_3}}{4 M_{S_3}^2|C_{VLL}^{\rm SM}|\sin\theta_{ct}}\right)^2 \lsim 26\, .
\end{eqnarray}
The bounds from mixing and from direct searches are the same as in the case of the model of Section~\ref{sec:LQS1}, implying that $R(K_S\to\mu^+\mu^-)_{\ell=0}\approx 26$ can be saturated. The interference term in this case could thus be enhanced compared to the SM by a factor of ${\cal O}(5)$.

\subsection{Scalar doublet (2HDM)}
%
Another example of a simple model that can contribute to $K_S\to(\mu^+\mu^-)_{\ell=0}$ is a two-Higgs-doublet model (2HDM), in which a second scalar doublet is added to the SM,
\begin{equation}
    \Phi \, \sim \, (1,2)_{\frac{1}{2}} \, = \begin{pmatrix} \phi^+ \\ \phi_0 \end{pmatrix}\, .
\end{equation}
If $\phi_0$ couples to either $(\bar s_L d_R)$ or $(\bar d_L s_R)$, and to $(\bar\mu_L \mu_R)$, it would contribute to $K_S\to(\mu^+\mu^-)_{\ell=0}$.
Without loss of generality, we choose to align the neutral state with the down-type mass eigenstates, The relevant Lagrangian terms are then,
\begin{eqnarray}\label{eq:LagPhi}
    {\cal L}_{\Phi} \supset  \lambda^d_{ij}\Bigg[ \phi_0 (\bar d_L)_i (d_R)_j  + \phi^+ (\bar u_L)_k V_{ki} (d_R)_j  + h.c.\Bigg] + \lambda_{22}^e\Bigg[\phi_0 \bar\mu_L \mu_R  + \phi^+ \bar{\nu_\mu}_L \mu_R + h.c.\Bigg]\, ,
\end{eqnarray}
with $(i,j) = (1,2), (2,1)$.
After integrating out the $\Phi$ fields, the effective dimension six operators $O_{SLR}, O_{SRL}$ are generated, with coefficients
\begin{eqnarray}
    C_{SLR}^\Phi = \frac{\lambda_{21}^d {\lambda_{22}^e}^*}{ M_\phi^2}\, , \qquad  C_{SRL}^\Phi = \frac{\lambda_{12}^d {\lambda_{22}^e}^*}{ M_\phi^2}\, .
\end{eqnarray}
The contribution to $R(K_S\to\mu^+\mu^-)_{\ell=0}$ is given by Eq.~(\ref{eq:generalKS}),
\begin{eqnarray}
	R(K_S\to\mu^+\mu^-)_{\ell=0}^{\Phi} \, = \, \Bigg(1 + \frac{m_K^2 / m_\mu}{(m_s+m_d)}\frac{(|\lambda_{21}^d{\lambda_{22}^e}|\sin\Theta_{\phi_{21}}+|\lambda^d_{12}{\lambda_{22}^e}|\sin\Theta_{\phi_{12}})}{M_\phi^2 |C^{\rm SM}_{VLL}|\sin\theta_{ct}}\Bigg)^2 \, ,
\end{eqnarray}
where 
\begin{eqnarray}
    \Theta_{\phi_{21(12)}}  \equiv \half\arg\left(\frac{q}{p}\right)-\arg(\lambda_{21(12)}^d{\lambda_{22}^e}^*)\, .
\end{eqnarray}
In order to saturate the current upper bound, $R(K_S\to\mu^+\mu^-)_{\ell=0} \approx 1280$, we require
\begin{eqnarray}\label{eq:phiSaturate}
    |\lambda_{22}^e|(|\lambda_{21}^d|\sin\Theta_{\phi_{21}} + |\lambda^d_{12}|\sin\Theta_{\phi_{12}}) \approx \, 3 \cdot 10^{-5} \,\left(\frac{M_\phi}{\rm TeV}\right)^{2}\, .
\end{eqnarray}
This model induces $K^0 - \overline K{}^0$ mixing at tree level. We find the following constraints on the magnitude of couplings~\cite{UTfit:2007eik},
\begin{eqnarray}\label{eq:phiMixing}
    |\lambda^d_{12}|^2,\,|\lambda^d_{21}|^2 &\lsim & 10^{-8} \left(\frac{M_\phi}{\rm TeV}\right)^2 \, , \\ \nonumber
    |\lambda^d_{12}\lambda^d_{21}| & \lsim & 10^{-9} \left(\frac{M_\phi}{\rm TeV}\right)^2 \, .
\end{eqnarray}
From Eq.~(\ref{eq:phiSaturate}), and assuming a perturbative coupling, $|\lambda^e_{22}|\lsim 1$, we deduce that in order to saturate the bound we need at least one of the couplings $\lambda^d_{12},\lambda^d_{21}$ to obey
\begin{equation}
    |\lambda^d_{ij}|\geq |\lambda^e_{22}\lambda^d_{ij}|\sin\Theta_{\phi_{ij}} \approx 3\cdot 10^{-5} \left(\frac{M_\phi}{\rm TeV}\right)^2\, ,
\end{equation}
which can be accommodated together with the constraints of Eq.~(\ref{eq:phiMixing}), independently of $M_\phi$.
Loop diagrams involving a muon loop in principle constrain the phases $\Theta_{\phi_{12}},\,\Theta_{\phi_{21}}$, however, these are strongly suppressed and do not result in relevant bounds.

Models that induce scalar dimension six operators are prone to break the assumption of $|A(K_L)_{\ell=1}|=0$, since they introduce, in general, both pseudo-scalar (CP-odd) and scalar (CP-even) leptonic currents. We therefore comment on two scenarios:
\begin{itemize}
	\item Models with $\lambda^d_{12} = \lambda^d_{21}$ : \\
	In this scenario we have $C_{SLR} = C_{SRL}$, which implies that only the CP-odd leptonic current $(\bar\mu\gamma^5\mu)$ is generated, and no contribution exists for the $(\mu^+\mu^-)_{\ell=1}$ final state, so that the assumption of $|A(K_L)_{\ell=1}|=0$ is fulfilled. 
	This ensures that the clean short-distance parameter $|A(K_S)_{\ell=0}|$ can be extracted from the time dependent $K\to\mu^+\mu^-$ rate.

	\item Models with $\lambda_{12}^d \neq \lambda_{21}^d$ :\\
	If no symmetry protects $\lambda^d_{12} = \lambda^d_{21}$, this model generally induces $|A(K_L)_{\ell=1}| \neq 0$ (unless ${\cal R}e(\lambda_{12}^d\lambda^e_{22}\left(p/q\right)^{1/2}),\,{\cal R}e(\lambda_{21}^d\lambda^e_{22}\left(p/q\right)^{1/2})=0$, that is, $\cos\Theta_{\phi_{12}},\, \cos\Theta_{\phi_{21}} = 0$). This breaks the assumption needed in order to extract short-distance parameters from the measurement of the interference terms.  The observable of Eq.~(\ref{eq:extractSD}) is no longer a pure measurement of a short-distance parameter, but is polluted by irreducible long-distance effects. 
	
\end{itemize}

We conclude that models with a second scalar doublet can significantly enhance the ${\cal B}(K_S\to\mu^+\mu^-)_{\ell=0}$ rate, saturating the current experimental bound. If no symmetry protects the relation $\lambda^d_{12} = \lambda^d_{21}$, these models will generally lead to non-zero $|A(K_L)_{\ell=1}|$, which extinguishes the ability to extract ${\cal B}(K_S\to\mu^+\mu^-)_{\ell=0}$ from a measurement of $\Gamma_{int.}$.
The total rate, ${\cal B}(K_S\to\mu^+\mu^-)$, could still exhibit significant enhancement compared to the SM, signaling NP is at play.

\section{Discussion and Conclusion}
\label{sec:conclusion}
%
Following the recent understanding that short-distance parameters of the SM can be cleanly extracted from a measurement of interference effects in $K\to\mu^+\mu^-$~\cite{Dery:2021mct,DAmbrosio:2017klp}, we have addressed the question of what can be learned from such a measurement beyond the SM.
Any NP contribution in which the leptonic current is CP-odd, as is the case to a good approximation within the SM, keeps effects of CPV limited to a single partial wave configuration, $(\mu^+\mu^-)_{\ell=0}$, which enables the extraction of the purely short-distance observable, ${\cal B}(K_S\to\mu^+\mu^-)_{\ell=0}$.
NP that induces also CP-even leptonic currents, as is the case in general when scalar operators are induced, can also result in large enhancements to ${\cal B}(K_S\to\mu^+\mu^-)_{\ell=0}$, but would not allow its clean extraction from interference effects.

The current model-independent bound on ${\cal B}(K_S\to\mu^+\mu^-)_{\ell=0}$ is determined by the LHCb bound on the total branching ratio and given by 
\begin{equation}
    R(K_S\to\mu^+\mu^-)_{\ell=0} \lsim 1280\, ,
\end{equation} 
leaving room for enhancement of up to ${\cal O}(30)$ compared to the SM, at the amplitude level.
In the future, LHCb is expected to improve its reach by an order of magnitude, allowing to probe amplitude enhancements of ${\cal O}(10)$ times the SM contribution.

We note, however, that a measurement of the total branching ratio is a measurement of the sum, ${\cal B}(K_S\to\mu^+\mu^-)_{\ell=0} + {\cal B}(K_S\to\mu^+\mu^-)_{\ell=1}$. Therefore, while it can probe the existence of large NP contributions, it cannot allow the extraction of short-distance parameters.
A dedicated measurement of $K_S - K_L$ interference effects in $K\to\mu^+\mu^-$ is required in order to obtain a clean evaluation of the pure short-distance quantity, ${\cal B}(K_S\to\mu^+\mu^-)_{\ell=0}$.  

We have formulated relations between $K_S\to(\mu^+\mu^-)_{\ell=0}$ and $K_L\to\pi^0\bar\nu\nu$ within several EFT scenarios. We find that within models with lepton-flavor universality in which a single vectorial dimension six operator is present, involving the lepton doublet, the two modes are correlated. However, if scalar operators are at play, or if more than one vectorial operator is present, the two modes are independent. 
CPV in additional modes, such as $K_L\to\pi^0\mu^+\mu^-$ and $K_S\to\pi^0\pi^0\mu^+\mu^-$, is expected to have analogous relations. We leave the study of these relations to a future work. 

Within specific NP models, constraints from additional observables are relevant, arising from $K^0-\overline K{}^0$ mixing and from direct searches for NP resonances. 
We have presented examples of simple explicit models in which large enhancements to ${\cal B}(K_S\to\mu^+\mu^-)_{\ell=0}$ are possible. 
We find that models in which a scalar leptoquark is added to the SM, as well as a 2HDM, can allow large enhancements without violating existing constraints.
Of the three models, two can saturate the current experimental bound on ${\cal B}(K_S\to\mu^+\mu^-)_{\ell=0}$, while the third is an example where a relation to ${\cal B}(K_L\to\pi^0\bar\nu\nu)$ implies a constraint coming from the GN bound. 

The models we consider, while non-generic, affect the kaon sector alone, and cannot be probed by measurements in other sectors. 
This provides additional motivation for dedicated kaon programs in next-generation experiments.
Initial estimates of the feasibility of reaching SM sensitivity in a measurement of interference effects in $K\to\mu^+\mu^-$ in next-generation experiments are very encouraging~\cite{Dery:2021mct}. Our results indicate that such a measurement would be a unique and potent probe of physics beyond the SM.

\begin{acknowledgments}
    We thank Yuval Grossman for many useful discussions. We thank Yossi Nir and Stefan Schacht for helpful comments on the manuscript. 
\end{acknowledgments}


\end{document}